\documentclass[12pt]{article}

\newcommand{\br}{\hskip .25cm/\hskip -.25cm}
\begin{document}

\sloppy
\begin{flushright}{SIT-HEP/TM-6}
\end{flushright}
\vskip 1.5 truecm
\centerline{\large{\bf Enhanced baryon number violation}}
\centerline{\large{\bf due to cosmological defects }}
\centerline{\large{\bf with localized fermions along extra dimension}}
\vskip .75 truecm
\centerline{\bf Tomohiro Matsuda
\footnote{matsuda@sit.ac.jp}}
\vskip .4 truecm
\centerline {\it Laboratory of Physics, Saitama Institute of
 Technology,}
\centerline {\it Fusaiji, Okabe-machi, Saitama 369-0293, 
Japan}
\vskip 1. truecm
\makeatletter
\@addtoreset{equation}{section}
\def\theequation{\thesection.\arabic{equation}}
\makeatother
\vskip 1. truecm

\begin{abstract}
\hspace*{\parindent}
We propose a new scenario of baryon number violation in models with 
extra dimensions.
In the true vacuum, baryon number is almost conserved due to the
 localization mechanism of matter fields, which 
suppresses the interactions between quarks and leptons.
We consider several types of cosmological defects in
four-dimensional spacetime that shift the center of the localized
matter fields, and show that the magnitudes of the baryon number violating
interactions are well enhanced.
Application to baryogenesis is also discussed.
\end{abstract}

\newpage
\section{Introduction}
\hspace*{\parindent}
Although the quantum field theory made a great success, there is no
consistent scenario in which the quantum gravity is included.
The most promising scenario in this direction would be the string
theory, where the consistency is maintained by the requirement of
additional dimensions.
At first the sizes of extra dimensions had been assumed to be as small as
$M_p^{-1}$, however it has been observed later that there is no reason
to expect such a tiny compactification radius\cite{extra_1}.
Denoting the volume of the $n$-dimensional compact space by $V_n$,
the observed Planck mass is obtained by the relation $M_p^2=M_{*}^{n+2}V_n$,
where $M_{*}$ denotes the fundamental scale of gravity.
If we assume more than two extra dimensions, $M_*$ can be assumed to be
close to the electroweak scale without conflicting any observable
bounds.
Although such a low fundamental scale considerably improves the standard
situation of the hierarchy problem, the scenario requires some degrees of
fine-tuning.
The largeness of the quantity $V_n$ is perhaps the most obvious example
of such fine-tuning.
There are of course other aspects of fine-tuning, which are 
common to conventional scenarios of grand unified theories(GUT).
In particular, the compatibility between the stability of proton and
baryogenesis may be the most 
problematic in models of such a low fundamental mass scale.
In theories with low fundamental scale, the suppression can not be
achieved by merely increasing the mass scale, and some non-trivial
mechanism is needed.
There is an interesting mechanism suggested in ref.\cite{proton}, where
a dynamical mechanism for localization of fermions on the thick wall is
adopted to solve the problem of the proton stability.
In this scenario leptons and baryons are localized at displaced
positions in the extra space, where the smallness of their interaction is 
insured by the smallness of the overlap of their wavefunctions along the
extra dimension.
On the other hand, the observed baryon number asymmetry of the Universe
requires baryon number violating interactions to have been effective but
non-equilibrium in the early Universe.
In general the production of net baryon asymmetry requires baryon number
violating interactions, C and CP violation and a departure from the
thermal equilibrium\cite{sakharov}.
In the case that the fundamental mass scale is sufficiently high, 
the first two of these ingredients are naturally contained in
conventional GUTs or other string-motivated scenarios, and the 
third can be realized in an expanding universe where it is not uncommon that
interactions come in and out of equilibrium, producing the stable
heavy particles or cosmological defects.
In the original and simplest model of baryogenesis\cite{original, decayb},
a heavy GUT gauge or Higgs boson decays out of equilibrium producing
a net baryon asymmetry.
In our case, however, the situation is rather involved because of the
low fundamental mass and the resulting low reheat temperature,
which makes it much more difficult to produce the baryon asymmetry while
achieving the proton stability in the present Universe\cite{lowB}.
In this respect, it is very important to propose ideas to
enhance the baryon number violating interactions that can appear
even if the reheating temperature is low. 
In this letter we propose a mechanism where the enhancement of the baryon
number violating interaction is realized by the several types of the
cosmological defects that can survive well below the TeV scale.
Our key idea is that the five-dimensional mass of the fermions, which
determines the position of the wavefunctions of the fermions, can 
depend on the vacuum expectation value of some five-dimensional
scalar field.
We also assume that there are cosmological defect configurations of such
scalar fields. 
In this case, depending on their effective couplings to the inflaton,
cosmological defects can be formed at the (non-equilibrium) reheating
period of the inflaton. 
The positions of the localized fermions vary in the defect
configurations such as strings or monopoles, or in the quasi-degenerated
false vacuum of the corresponding domain wall.
Such a shift of the center of the wavefunction along the extra dimension
can make the tiny baryon number violating interactions enhanced to
produce the sufficient baryon number asymmetry.
We consider the case where the fermionic mass in
the five-dimensional theory, which had been assumed to be a constant in
the original model, depends on the vacuum expectation value of the
five-dimensional scalar field that is different from the one
constitutes the fat kink configuration along the extra dimension.
As we have discussed above, the most attractive effect is the
enhancement of the baryon number violating interactions that can be a
promising candidate to explain the baryogenesis with large extra
dimensions. 

\section{Defects and domains}
\hspace*{\parindent}
Localizing fields in the extra dimension necessitates breaking of higher
dimensional translation invariance, which is accomplished by a spatially
varying expectation value of the five-dimensional scalar field
$\phi_{A}$ of the thick wall along the extra dimension.\footnote{Here we
limit ourselves to constructions with fermions 
localized within only one extra dimension\cite{proton}.
Generalizations to higher dimensions are straightforward, which 
is already discussed in ref.\cite{higher}.}
If the scalar field $\phi_{A}$ couples to the five-dimensional 
fermionic field $\psi$ through the five-dimensional Yukawa interaction
$g \phi_A \overline{\psi}\psi$, whose expectation value $<\phi_A>$ varies 
along the extra dimension but is constant on four-dimensional world,
it is possible to show that the fermionic field localizes at the place
where the total mass in the five-dimensional theory vanishes.
For definiteness, we consider the Lagrangian
\begin{eqnarray}
{\cal L} &=&\overline{\psi_{i}}\left(i \br{\partial_5} +g_{i}\phi_{A}(y) 
+m_{5,i}
\right)\psi_{i}\nonumber\\
&&+\frac{1}{2}\partial_{\nu}\phi_{A} \partial^{\nu}\phi_{A} \nonumber\\
&&-V(\phi_{A}),
\end{eqnarray}
where $y$ is the fifth coordinate of the extra dimension.
For the special choice $\phi_{A}(y)=2\mu^2 y$, which corresponds to 
approximating the kink with a straight line interpolating two vacua,
the wave function in the fifth coordinate becomes gaussian centered
around the zeros of $g_{i}\phi_{A}(y)+m_{5,i}$.
It is also shown\cite{extra_1} that a left handed chiral fermionic field
in the four-dimensional representation can result from the localization
mechanism.
The right handed part remains instead delocalized in the fifth
dimension.
The above idea can be utilized to certificate the proton stability.
When leptons and baryons have the five-dimensional masses
$m_{5,l}$ and $m_{5,q}$, the corresponding localizations are
at $y_l=-\frac{m_{5,l}}{2g_l \mu^2}$ and $y_q=-\frac{m_{5,q}}{2g_q
\mu^2}$, respectively.
Even if the five-dimensional theory violates both baryon and lepton
number maximally, the dangerous operator in the effective
four-dimensional theory is safely suppressed.
For example, we can expect the following dangerous operator in the
five-dimensional theory,
\begin{equation}
{\cal O}_5\sim \int d^5 x \frac{QQQL}{M_*^3}
\end{equation}
where $M_*$ denotes the fundamental mass scale and $Q,L$ are the
five-dimensional representation of the fermionic field.
The corresponding four-dimensional proton decay operator is obtained by
simply replacing the five-dimensional fields by the zero mode fields and
calculating the wave function overlap along the fifth dimension $y$.
The result is
\begin{equation}
{\cal O}_4 \sim \epsilon \times \int d^4 x \frac{qqql}{M_{*}^2},
\end{equation}
where $q,l$ denotes the four-dimensional representation of the chiral
fermionic field.
The overlap of the fermionic wavefunction along the fifth dimension is
included in $\epsilon$.
For a separation $r=|y_b - y_l|$ of $\mu r =10$, one can obtain
$\epsilon \sim 10^{-33}$ which makes this operator safe even for 
$M_*\sim$TeV.

Let us extend the above idea to include another scalar field 
$\phi_B$
that determines the five-dimensional mass $m_5$ as well as the position
of the center of the fermionic wavefunction along the fifth 
dimension.\footnote{
We can utilize the idea of the orbifold boundary conditions
that produce 
chiral fermion zero modes in compactified higher dimensional theories
and provides a simple and explicit realization of the separation of
quarks and leptons in the fifth dimension\cite{orbifold}.
As is discussed in ref.\cite{orbifold}, one can obtain the localized
fermions at the fixed points at $y=0$ or $y=L$.
In this case one can obtain two degenerated solutions,
one is the positive configuration for $0<y<L$, and the other is
the negative one\cite{orbifold}.
If the sign is positive, the zero-mode is concentrated at $y=0$.
If it is negative, the zero mode is concentrated at $y=L$.
In general, two degenerated vacua generates the domain configuration,
separated by the domain wall interpolating between them.
If the splittings of the baryons and the leptons are induced by the
above-mentioned mechanism of the orbifold, and if one of the scalar
field develops domain wall structure, splitting can be  dissolved in
the quasi-degenerated false vacuum.
Then the baryon number violation is maximally enhanced in the false
vacuum, which helps the scenario of baryogenesis by the decaying heavy X
bosons.}
We assume that the additional scalar field does {\it not} make kink
configuration along the fifth dimension, but {\it does} make a defect
configuration in the conventional four-dimensional space. 
For definiteness, we consider the $\phi_B$-dependent five-dimensional
mass $m(\phi_B)_{5,i}$.
$\phi_{A}$ makes the kink configuration along the fifth dimension
while $\phi_B$ develops defect configuration in the four-dimensional
spacetime.
Here we consider the simplest case where $m_{5,i}$ are given by
$m(\phi_{B})_{5,i}=k_{i} \phi_{B}$, and the potential for $\phi_B$
is given by the double-well potential of the form; 
$V_{B}=-m_B \phi_{B}^2 +\lambda_B \phi_B^4$.
In our simplest example, because of the effective $Z_2$ symmetry of the
scalar field $\phi_B$, the resultant defect is the cosmological domain
wall.
Of course the effective $Z_2$ symmetry can be explicitly broken by the
gravitational effect or the higher-dimensional operators suppressed by
the cut-off scale\cite{vilenkin, matsuda_wall}. 
One can easily extend the model to include the string or monopole
configuration in four-dimensional spacetime, if the appropriate symmetry
is imposed on the scalar field $\phi_B$.
The most obvious example is the choices of the form,
\begin{equation}
m(\phi_B)_{5,i}=k_{i}\frac{|\phi_{B}|^2}{M_*}
\end{equation}
where $\phi_B$ is charged with $U(1)$.
In any case, the position of the fermionic wavefunction along the fifth
dimension can be modified by the defect configuration in the
four-dimensional spacetime.
The largest contribution is expected in the quasi-degenerated vacuum
of the cosmological domain-wall that interpolates between $\phi_B = \pm
v$.
Let us assume that the wavefunctions of the quarks and the leptons are
localized at the opposite side of the $\phi_A$ kink along the fifth
dimension so that their wavefunctions are well separated.
We also assume for simplicity that the five-dimensional masses $m_{5}$ for
the leptons are constant. 
In the false vacuum domain of the $\phi_B$ wall, the centers of the quark 
wavefunctions move toward the opposite side of the $\phi_A$ wall.
In this case the distances between quarks and leptons are changed by 
O(1), which drastically modifies the magnitude of the baryon number
violating interactions in the false vacuum.

Although it seems rather difficult to produce these defects merely by the
thermal effect after inflation, nonthermal effect may create such
defects during reheating period of inflation.
Nonthermal creation of matter and defects has raised a remarkable
interest in the last years.
In particular, efficient production of such products during the period
of coherent oscillations of the inflaton has been studied by many
authors\cite{PR}.
In this letter, however, we will not go into the details of such processes 
but simply assume the situation that the defects are efficiently
produced after inflation by thermal or nonthermal effects.\footnote{
There is a possibility that the defects are generated after the
first brane inflation, while the reheat temperature after the second thermal
brane inflation is kept much lower than the electroweak
scale\cite{thermalbrane}. 
The cosmological constraint on the domain wall that is produced before
thermal inflation is already discussed in ref.\cite{matsuda_wall}.}
Thermal effects may become more important when one considers the
supersymmetric models where the positions of the localized matter fields
may be parameterized by flat directions.
In such cases, one can expect thermal symmetry restoration at the
temperature much lower than the cut-off scale, which is accessible in
realistic scenarios.
During thermal symmetry restoration, if the five-dimensional mass terms
$m_{5,i}$ are all determined by a field $\phi_B$ that parameterizes the flat
direction, both leptons and baryons may be localized at $y=0$.
During this period the baryon number violation becomes maximal and the
baryogenesis by the decaying heavy $X$ bosons can be promising.
We will discuss this issue for supersymmetric grand unified theories 
in the forthcoming paper\cite{matsuda_tobe}.

\section{Enhanced baryon number violation and baryogenesis}
\hspace*{\parindent}
In this section we explore the possibility of obtaining sufficient
baryon number asymmetry of the Universe in models with localized
fermions.
Here we focus our attention to the baryogenesis by the decaying heavy
particles.

First we consider a simplest model of baryogenesis\cite{decayb} with two
species of heavy bosons $X_{i}$ which can decay into quarks and leptons,
through the effective four-dimensional interactions of the form;
\begin{eqnarray}
{\cal L}_{Xqq}&=& \lambda_{1}X\overline{q}\overline{q}\nonumber\\
{\cal L}_{Xlq}&=& \lambda_{2}Xlq,
\end{eqnarray}
where $\lambda_2$ contains the tiny suppression factor $\epsilon_2$ of
the form $\epsilon_2\sim e^{-\mu^2 r^2}$.
The baryogenesis with the enhanced baryon number violating interactions
is already discussed in ref.\cite{thermalB}, where they have expected
that the thermal effect modifies the suppression factor, and
concluded that the baryogenesis 
mediated by the heavy bosons becomes successful if the suppression
factor $\epsilon_2$ is enhanced to be larger than $e^{-40}$.
Unfortunately, the thermal effect is so weak in generic situations 
that the enhancement is not enough to produce the realistic
baryon number of the Universe.

Now we consider the case that the five-dimensional mass depends on an another
scalar field $\phi_B$ and the vacuum expectation value of $\phi_B$ is
determined by the effective $Z_2$-symmetric potential.\footnote{
In general the effective $Z_2$ symmetry of the effective theory can be
broken explicitly by the ultraviolet interactions.
The explicit-breaking operators can appear in the effective Lagrangian
suppressed by the cut-off scale, and destabilize the domain wall
configuration. 
The condition for the safe decay is discussed in ref.\cite{vilenkin}} 
Expecting the generic double-well potential for $\phi_B$, there should
be quasi-degenerated vacua where $\phi_B$ changes its sign.
Let us imagine the situation that the quarks and leptons are placed at
the opposite side of the $\phi_A$ kink, and their large distance
certificates the proton stability in the true vacuum.
On the other hand, when one of the five-dimensional mass changes its
sign in the false vacuum, the distance can be modified by the factor of
O(1).
For example, if the distance $r$ becomes $\frac{1}{2} r$ in the false vacuum,
the suppression factor $10^{-16}$ becomes $10^{-4}$, which is about
$10^{12}$ times larger than the conventional value.
This effect is obviously enough to explain the baryon number asymmetry
in the scenario of ref.\cite{thermalB}.
Although it seems easy to reduce the amount of the baryon number asymmetry
in this scenario, it depends on the details of the reheating process
of the inflation, which is beyond the scope of this letter.

Of course one can consider other cosmological defects, such as strings or
monopoles.
In generic situations, monopoles are not effective for baryogenesis,
because of their small volume factor.
However, in some realistic cases, strings can become effective source of
baryon number violation that may either washout or produce the baryon
asymmetry of the Universe\cite{stringbaryo}.
It is easy to introduce string configurations in our model.
The most promising way would be to consider the vortex solution along
the extra dimensions in place of the wall-like kink 
configuration\cite{vortex}.
However, one can include the string configuration without modifying the
simplest situation of the original idea.
For example, we can consider the mass term:
$m(\phi_B)_{5,i}=\lambda_{i}\frac{|\phi_B|^2}{M_*}$.
In this case, the most interesting situation is that the position of
both leptons and baryons are all determined by $\phi_B$,
and their distance depends on their coupling constants.
In the core of the string where $\phi_B$ vanishes, leptons and baryons
are centered at the same point along the fifth dimension.
The suppression disappears in the string, which is similar to the
scenarios of the string-mediated baryon number violation in the 
conventional GUTs.
In our model, however, we are not assuming
the GUT-like symmetry restoration inside the string.
Around the string, both the scattering and their decay by the loop 
can become the effective source of the baryon number violation.
Although the idea of string-mediated baryogenesis is very attractive,
it contains many kinds of models to be discussed.
Detailed studies in this direction are given in the forthcoming
paper\cite{matsuda_tobe} to avoid comlexities.

\section{Conclusions and Discussions}
\hspace*{\parindent}
In this letter we have proposed a new scenario of baryon number violation
in theories with localized fermion wavefunctions along the extra
dimension.
The baryon number can be almost conserved in the true vacuum by the
localization mechanism, while it is well enhanced in the
background of cosmological defect configurations.
The baryon number violating interactions are most effective for the
cosmological domain walls, where the domain of the false vacuum appears.
It is convincing that this interesting idea opens new possibilities
for baryogenesis with extra dimensions, which is not discussed in the past.

\section{Acknowledgment}
We wish to thank K.Shima for encouragement, and our colleagues in
Tokyo University for their kind hospitality.

\end{document}